\documentclass[twocolumn,showpacs,superscriptaddress,amsmath,amssymb]{revtex4}
\usepackage{graphicx}
\usepackage{float}
\usepackage{color}
\usepackage{amsmath}
\usepackage{booktabs}
\usepackage{float}

\newcommand{\overbar}[1]{\mkern 1.5mu\overline{\mkern-1.5mu#1\mkern-1.5mu}\mkern 1.5mu}
\usepackage{verbatim}

\begin{document}

\title{Measuring Nonlinear Stresses Generated by Defects in 3D Colloidal Crystals}

\author{Neil Y.C. Lin}
\affiliation{Department of Physics, Cornell University, Ithaca, NY 14853}
\author{Matthew Bierbaum}
\affiliation{Department of Physics, Cornell University, Ithaca, NY 14853}
\author{Peter Schall}
\affiliation{Institute of Physics, University of Amsterdam, Amsterdam, The Netherlands}
\author{James P. Sethna}
\affiliation{Department of Physics, Cornell University, Ithaca, NY 14853}
\author{Itai Cohen}
\affiliation{Department of Physics, Cornell University, Ithaca, NY 14853}
\date{\today}

\begin{abstract} 
The mechanical, structural, and functional properties of crystals are determined by their defects~\cite{schiotz1998softening, peng2010melting, alsayed2005premelting, hull1984introduction}. The distribution of stresses surrounding these defects have broad implications for our understanding of transport phenomena including growth of voids~\cite{le1981model}, impurity diffusion~\cite{cowern1991experiments}, as well as dislocation creep and climb~\cite{hull1984introduction, schall2006visualizing}. When the defect density rises to levels routinely found in real-world materials, transport is governed by local stresses that are predominantly nonlinear~\cite{schiotz1998softening, schiotz2003maximum, li2002atomistic, cai2006non, lechner2009defect}. Such stress fields however, cannot be measured using conventional bulk and local measurement techniques. Here, we present the first direct and spatially resolved experimental measurements of the nonlinear stresses surrounding colloidal crystalline defect cores. Our measurements show that the nonlinear stresses at vacancy cores generate attractive interactions between them. In addition, we directly visualize the softening of crystalline regions surrounding dislocation cores, and find that stress fluctuations in quiescent polycrystals are uniformly distributed rather than localized to grain boundaries as is the case in strained atomic polycrystals. More broadly, these nonlinear stress measurements have important implications for strain hardening~\cite{bulatov2006dislocation}, yield~\cite{schiotz1998softening, schiotz2003maximum}, and fatigue~\cite{stephens2000metal}.   
\end{abstract}


\maketitle
Bulk measurements of the nonlinear materials response have shown that fascinating mechanical behaviors emerge when crystals are plastically deformed~\cite{hull1984introduction}. Such measurements however, average over the rich spatial heterogeneity in structure and stress distributions. This averaging makes it difficult to determine how microscopic mechanisms collude to determine a crystal's bulk behavior. Pioneering measurements of local crystalline strains have done much to elucidate the heterogeneity in the linear stress response of crystals~\cite{huang2013imaging, schall2006visualizing, bausch2003grain, irvine2010pleats, king2008observations, levine2006x}.  Despite these advances however, applying such techniques to measure the nonlinear stress distributions in crystals with defects has remained prohibitive since it is impossible to \textit{a priori} determine how the nonlinear modulus varies with strain or even define a strain when the structure is highly distorted. Consequently, it has been difficult to experimentally determine even the qualitative interactions between defects that give rise to these fascinating mechanical behaviors under large deformations. 

\begin{figure} [htp]
\includegraphics[height=0.25 \textwidth]{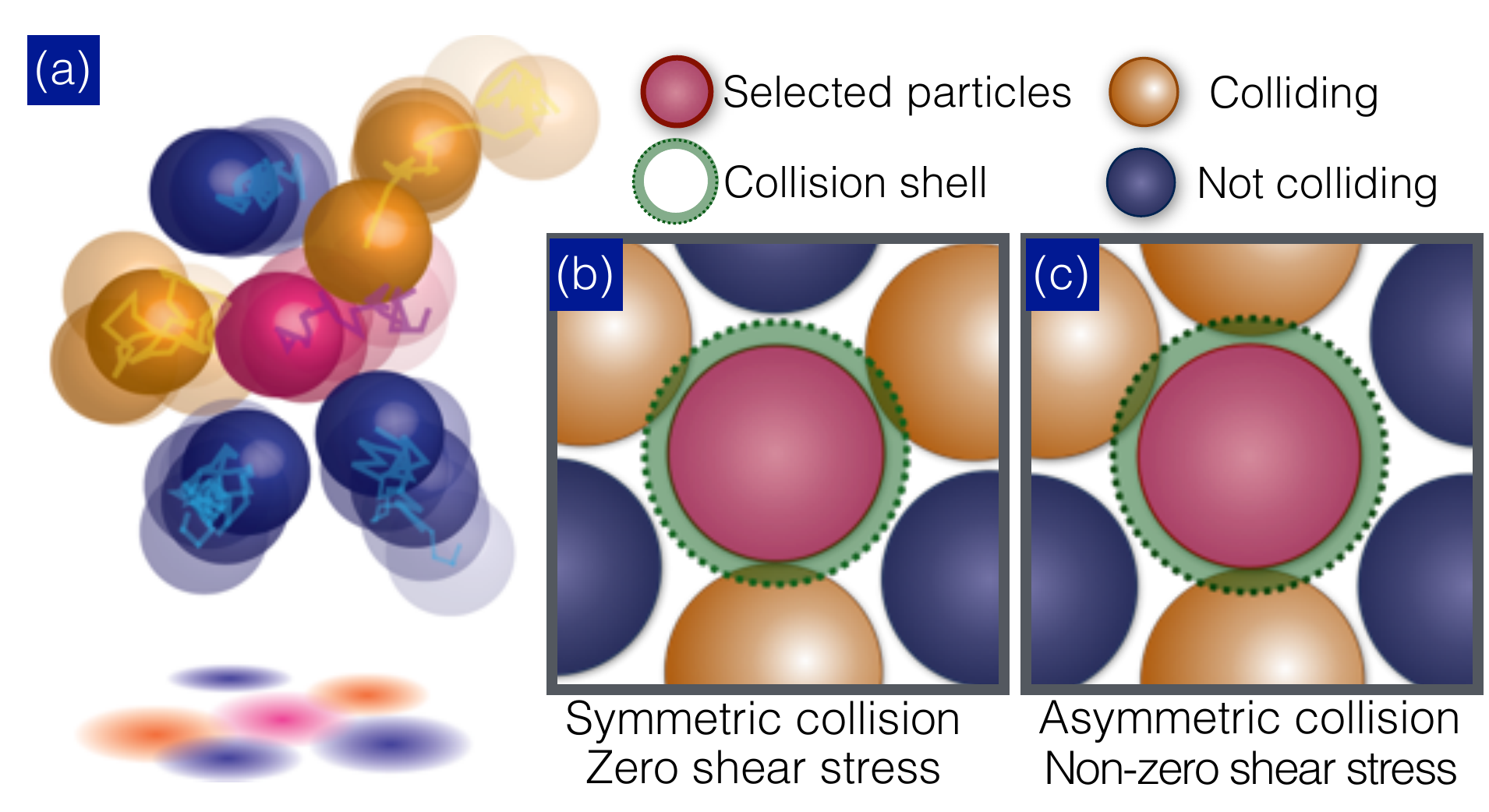}
\caption{\textbf{Particle-level stress measurements (SALSA).} (a) Particles exhibit Brownian motions (trajectories segments) and exert stresses on the selected particle (red sphere) when the neighboring particles collide with it. The energy density (stress) per collision is $k_B T/\Omega^\alpha$. (b) Schematic illustrating the SALSA algorithm for hard spheres. A thin shell ($\Delta$=106 nm) is constructed to identify colliding particles (yellow spheres), which lie within distance $2a+\Delta$ from the selected particle. The shear stress is zero when the colliding particles' configuration is symmetric. When the collisions are asymmetric, the shear stress is non-zero. The schematics here are two dimensional, but all presented calculations are fully three dimensional.}
\label{fig:SALSA_demo}
\end{figure}

\begin{figure*}
\includegraphics[height=0.34 \textwidth]{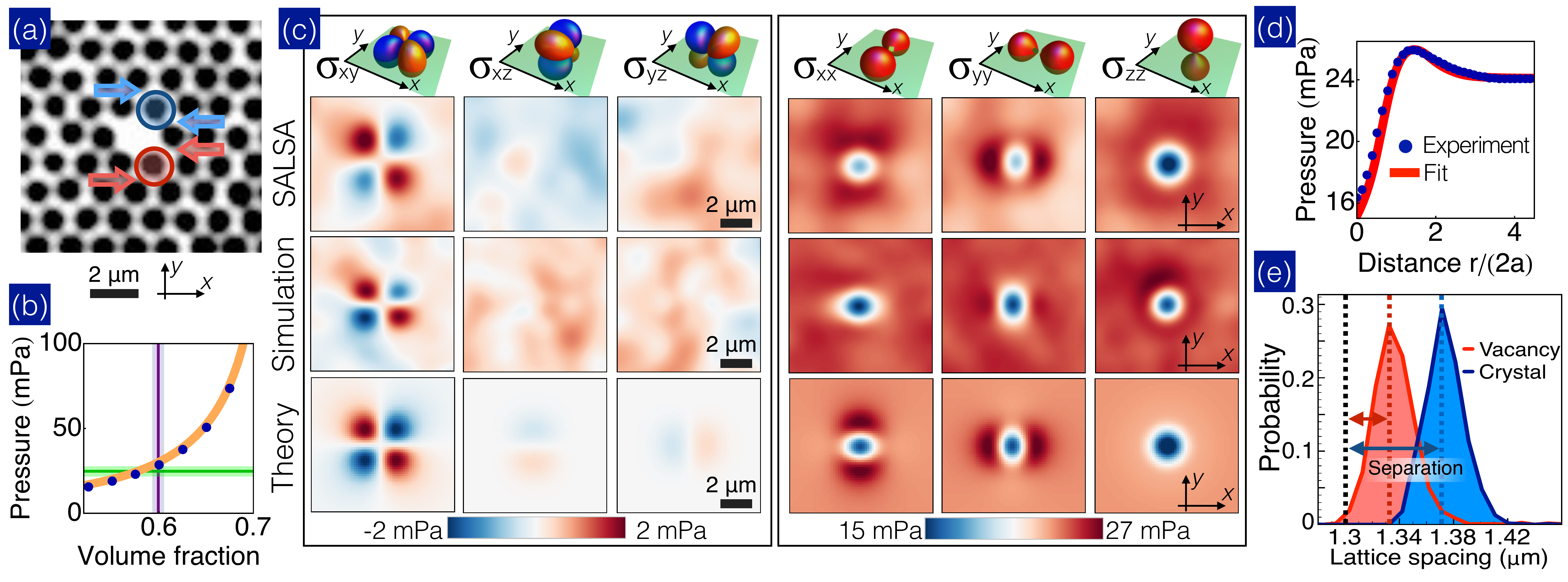}
\caption{\textbf{Stress around a vacancy.} (a) Confocal image of a crystal with an isolated vacancy defect with no other defects within five lattice spacings in the plane, or in the adjacent layers along the $(111)$ direction. The hard-sphere interparticle potential, large particle size, and high volume fraction slow down vacancy diffusion, which allows for time averaging. To account for particle polydispersity (SI) and the background stress we imaged 20 independent defects. Each vacancy is imaged for 20 seconds at a scan rate of 2 stacks per second, yielding a total of 800 snapshots for averaging. The absence of a particle at the defect core results in asymmetric collisions on the particles surrounding the vacancy. The upper right (blue) and lower right (red) particles are under positive and negative shear $\sigma_{xy}$. (b) SALSA accurately reports the mean pressure of the tested colloidal crystal with volume fraction 0.59 (purple line). Green line is the SALSA value. The blue dots are our hard sphere simulation results, and orange curve is the prediction from the literature. The shades of green and purple lines are the standard deviations. (c) All stress components around a vacancy determined using SALSA (upper row), simulation (middle row), and nonlinear elasticity (lower row). In contrast to the significant feature in $\sigma_{xy}$, we find that the small fluctuations in $\sigma_{xz}$ and $\sigma_{yz}$ are less than 20$\%$ of the variation in $\sigma_{xy}$. These measurements are consistent with the fact that both $\sigma_{xz}$ and $\sigma_{yz}$ exhibit nodes along the $x$-$y$ plane. Additionally, we find that the normal stresses $\sigma_{xx}$ and $\sigma_{yy}$ demonstrate elastic dipoles that align horizontally and vertically, respectively.(d) Pressure is plotted as a function of distance $d$ from the vacancy core. Both experimental (Blue) and theory (Red) results show clear stress enhancements at $d\sim3 a$. (e) Histograms of the particle separations near (red) and far away from (blue) the vacancy. The observed 50\% change in surface-surface spacing would correspond to a $\sim8.3\%$ local volume change in a defect free crystal. }
\label{fig:vacancy}
\end{figure*}

Here, building on the technological advances offered by high-speed confocal microscopy, we use Stress Assessment from Local Structural Anisotropy (SALSA) to directly measure the complete stress tensor down to the single particle-scale in a 3D colloidal crystal. Hard-sphere colloidal crystals have been widely employed as a model system to investigate many fundamental and important processes including defect nucleation~\cite{schall2006visualizing}, crystal melting~\cite{alsayed2005premelting, peng2010melting}, and crystal growth~\cite{van1997template}. In Brownian hard-sphere systems, the force with which particles collide can be related to the thermal energy $k_B T$. Therefore, using a time series of featured particle positions ~\cite{dinsmore2001three, crocker1996methods}, we determine the thermal collision probability, and directly report the stress arising from these Brownian collisions. Our derivation (see Supplemental Information) shows the stress tensor $\sigma^{\alpha}_{ij} = \sigma_{ij}(\mathbf{x}^{\alpha})$ at particle $\alpha$ can be approximated by 
\begin{equation}
\sigma^{\alpha}_{ij} =  \frac{k_B T}{\Omega^\alpha}\left(\frac{a}{\Delta}\right)  \langle \psi^{\alpha}_{ij} (\Delta) \rangle
\end{equation}
where $k_B T$ is the thermal energy, $\Omega^{\alpha}$ is the volume occupied by the particle, $a$ is the particle radius, and $\Delta$ is the cutoff distance from contact (SI and SI video). Here $\langle \psi^{\alpha}_{ij}(\Delta) \rangle$ is the time-averaged {\em local structural anisotropy} or fabric tensor for the particle, $\langle \psi^{\alpha}_{ij}(\Delta) \rangle=\langle \sum_{\beta \in nn} \hat{r}^{\alpha\beta}_{i}\hat{r}^{\alpha\beta}_{j} \rangle$, where $nn$ is the set of particles that lie within a distance $2a+\Delta$ from particle $\alpha$, $ij$ are spatial indices, and $\hat{r}^{\alpha\beta}$ is the unit vector between particle $\alpha$ and particle $\beta$. In the {\em local structural anisotropy} calculation, the trace $\hat{r}^{\alpha\beta}_{i}\hat{r}^{\alpha\beta}_{i}$ is the total number of neighbors while the remaining components captures the anisotropy of the collisions~\cite{bi2011jamming}. The time averaged fabric tensor of each particle accurately captures the probability of thermally induced collisions arising from the spatial distribution of its neighbors (Fig.~\ref{fig:SALSA_demo}). Scaling the probability by the energy density per collision $k_B T/\Omega^\alpha$, we then determine the Cauchy stress at the selected particle's position. This capability enables us to measure the local stress distributions surrounding crystalline defects such as vacancies (0D), dislocations (1D), and grain boundaries (2D). 
 
{\bf\emph{Vacancies}} dominate mass transport in crystals by playing key roles in electromigration growth of voids in integrated circuit interconnects, impurity diffusion, and dislocation creep and climb. These processes are governed by the vacancy interaction arising from the stress field. Whether the stress field surrounding the core is linear or nonlinear directly determines the qualitative interaction between vacancies and influences our understandings of those processes. To measure the stress field using SALSA, we create a crystal of $2a=$1.3 $\mu$m diameter silica particles via sedimentation in an index matched water-glycerol mixture. We image the 3D microstructure of isolated vacancies (Fig.~\ref{fig:vacancy} (a)) and determine their stress fields.

The mean pressure of our crystal sample is $\sim24$ mPa (green line in Fig~\ref{fig:vacancy}((b))), which is consistent with previous numerical predictions (orange curve)~\cite{alder1968studies} and our Brownian dynamics simulations (blue dots) for hard-sphere crystals at $\phi\sim 0.59$ (purple line). The top row of images in Fig~\ref{fig:vacancy}(c) show the vacancy 3D stress isosurfaces predicted by linear elasticity. The six independent stress components determined by SALSA are shown in the next row of Fig~\ref{fig:vacancy}(c). For simplicity we show 2D cuts of each stress component along the (111) or $x$\nobreakdash--$y$ plane (green planes) centered at the vacancy core in the upper images. We also conduct Brownian dynamics simulations (see SI) and directly calculate particle stresses (second to last row of Fig~\ref{fig:vacancy}(c)). The simulation results give quantitatively similar features for all stress components. For example, as shown in the first column of Fig~\ref{fig:vacancy}(c), $\sigma_{xy}$ exhibits a quadrupole distribution, which arises from the asymmetric collisions due to the absence of a particle at the vacancy core (blue and red arrows in Fig~\ref{fig:vacancy}(a)). 

The vacancy stresses also show non-trivial trends in the radial pressure distribution as shown in Fig~\ref{fig:vacancy}(d) that are not captured by isotropic linear elasticity. In particular, while linear elasticity predicts a constant pressure outside the vacancy core, here we observe a pressure bump at $r\sim3a$ that results from a $\approx 50\%$ reduction in particle surface separation near the core (double arrows in Fig~\ref{fig:vacancy}(e)). In hard sphere systems this reduced separation, hence increased local collision rate, leads to an enhancement of the local modulus.

To account for this changing modulus, we develop an isotropic elastic model including all terms up to third-order with finite strain. Using the volume change ($\Delta V = 8.4\%$) estimated in experiments and literature values of the bulk ($K = 93\,\rm{mPa}$) and shear ($\mu = 92\,\rm{mPa}$) moduli for our system's volume fraction~\cite{pronk2003large}, we fit the pressure distribution by adjusting the three third-order isotropic elastic constants. We find the predicted stress distributions quantitatively reproduce all stress components (last row of Fig~\ref{fig:vacancy}(c)) as well as the radial pressure distribution (red line Fig.~\ref{fig:vacancy}(e)). Furthermore, the local modulus at the pressure ring region can be determined from the fitting. We find that the bulk modulus at that region more than doubles to 213 mPa. This drastically increased modulus is consistent with the value from numerical studies of bulk hard spheres~\cite{pronk2003large, alder1968studies} at the local interparticle spacing of the pressure ring region. Overall, the strongly enhanced local modulus indicates a significant hardening near the defect core.

While linear isotropic theory predicts no interaction between vacancies, our findings indicate vacancies attract within the length scale associated with the pressure bump, as was predicted by numerical studies~\cite{bennett1971studies, dasilva2007formation, lechner2009defect}. This attraction can be understood by noting that the volume change $\Delta V$ due to one vacancy is negative and so the $P \Delta V$ term in the elastic energy leads to a force that attracts that vacancy to the pressure ring of the other (see SI). Therefore, we estimate the elastic energy of the attraction $\sim 2.6$ $k_BT$ at $r\sim4a$. Since this attraction is several times larger than the thermal energy, it will significantly accelerate the aggregation of vacancies. In an atomic crystal, this large vacancy aggregate will form a void. For hard sphere crystals without attractive interactions, void formation is inhibited by large configurational entropies found at very low equilibrium defect density. At the vacancy densities in many experimental systems, however, voids form in equilibrium~\cite{bennett1971studies} and neighboring particles surrounding a void will `evaporate' into the void, filling it with liquid-state particles in local equilibrium with the surrounding crystal.

\begin{figure*} [tp]
\includegraphics[height=0.25 \textwidth]{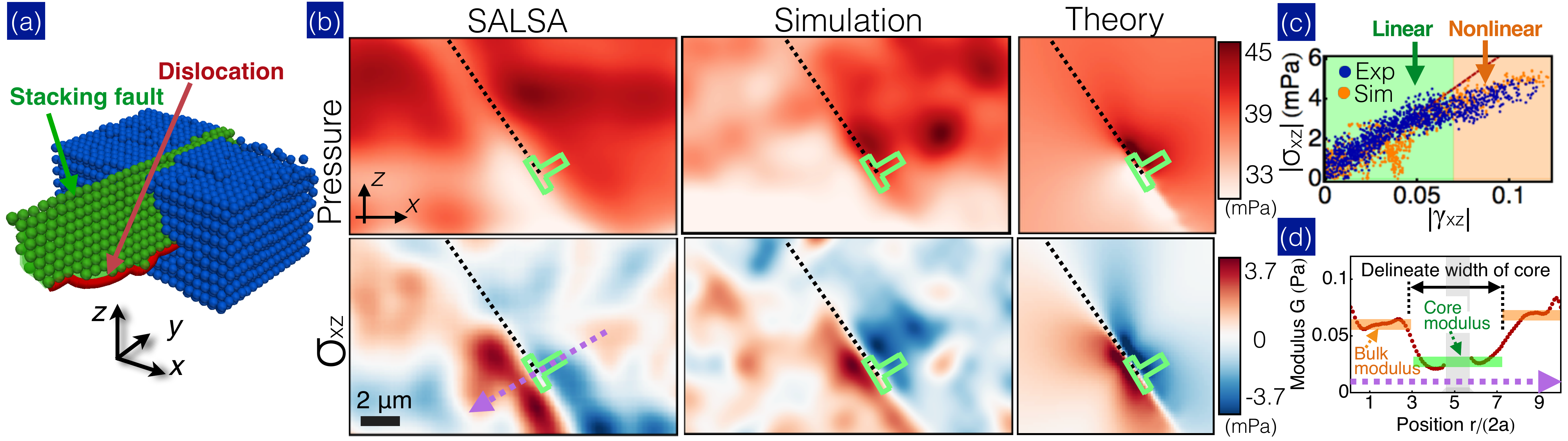}
\caption{\textbf{Dislocation Stress.} (a) 3D reconstruction of a partial dislocation (red line) and the associated stacking fault plane (blue) analyzed using the Dislocation Extraction Algorithm. (b) Pressure and shear stress, $\sigma_{xz}$, around a dislocation determined using SALSA (left column), simulation (middle column), and linear elasticity (right column). The experimental and simulation data are depth-averaged. In simulation, the experimental particle positions are used to determine appropriate initial and boundary conditions. The system is relaxed prior to recording the stresses to avoid particle overlaps due to featuring uncertainties. For the theory calculation, we use the observed Burgers vector and orientation of the partial dislocation to calculate the corresponding stress fields. (c) Both experimental and simulation stress-strain relations show softening behaviors at strains $|\gamma|\ge0.08$ (orange region). (d) Shear modulus versus position for fields within 2$\mu$m of the purple line in (b). The modulus decreases by $\sim$ 50\% at the defect core, which is approximately four particles wide. Since the modulus value fluctuates in the gray area due to the sign changes in stress and strain, the corresponding points are removed for clarity.}
\label{dislocation}
\end{figure*}

{\bf\emph{Dislocations}} are one-dimensional topological defects whose collective interactions determine macroscale plasticity including work hardening, yield stress, and fatigue. At the high defect densities involved in such processes however, interactions are significantly altered by nonlinear stress fields surrounding these defects. One critical conjecture that has been widely employed in the dislocation simulation literature is that the modulus softens at the dislocation core \cite{cai2006non, gracie2008new}. This conjecture however, has never been validated.

To study the dislocation stress field using SALSA, we grow a crystal on a patterned template with a lattice spacing 1.5$\%$ larger than the equilibrium crystal lattice. A 3D reconstruction of the particle configuration is shown in Fig.~\ref{dislocation}(a). The dislocation (red) delineates the lower bound of a stacking fault (green) embedded in a crystalline region (blue) which has been clipped for visual clarity. The dislocation is slightly curved (variation$\sim2a $) and aligned along the $y$-axis corresponding to the $(1\overbar{1}0)$ direction of the fcc lattice. The dislocation core is highlighted with a ($\boldsymbol{\bot}$) and has a Burgers vector $1/6(\overbar{1}\overbar{1}2)$, which corresponds to a Shockley partial, the most prominent dislocation in fcc metals.

Using SALSA we measure the stresses near the dislocation and show the pressure (upper row) and shear stress, $\sigma_{xz}$ (lower row) in Fig.~\ref{dislocation}. The stress field is averaged along the dislocation line to eliminate the effects of polydispersity. To confirm SALSA accurately extracts the stress features in this more complicated defect structure, we compare to stresses calculated by direct Brownian dynamics simulations that are seeded by the experiment data (middle column Fig.~\ref{dislocation}) (SI). Both experimental and simulation results show comparable features. Overall, we observe a pressure gradient across the stacking fault, and a shear stress dipole centered at the defect core. These general trends are consistent with predictions of linear isotropic elastic theory (right column Fig.~\ref{dislocation}) indicating that dislocation curvature does not qualitatively alter the stress distribution. However, both SALSA (blue) and the simulation (orange) results show a non-linear strain softening in highly strained regions near the defect core (Fig.~\ref{dislocation} (c)). This local modulus drop allows us to visualize the precise location and size of the dislocation core. To do so, we focus on the cross-section region denoted by the dashed line in Fig.~\ref{dislocation}(b), and plot nonlinear shear modulus ($d\sigma_{xz}/d\gamma_{xz}$) versus position ($r/2a$) in Fig.~\ref{dislocation}(d) (SI). The modulus decreases by $\sim50\%$ on both sides of the dislocation core, which is about four particles in width. Overall, our measured modulus profile clearly shows the softening and provides the first experimental evidence supporting the non-singular continuum assumption widely employed in dislocation theories and simulations~\cite{cai2006non, gracie2008new}, in which the divergence in the stress at the dislocation core is cut off. Moreover, this modulus softening regularizes the interactions between dislocations and dramatically influences the dislocation creep behavior in crystals.

\begin{figure} [tp]
\includegraphics[width=0.4 \textwidth]{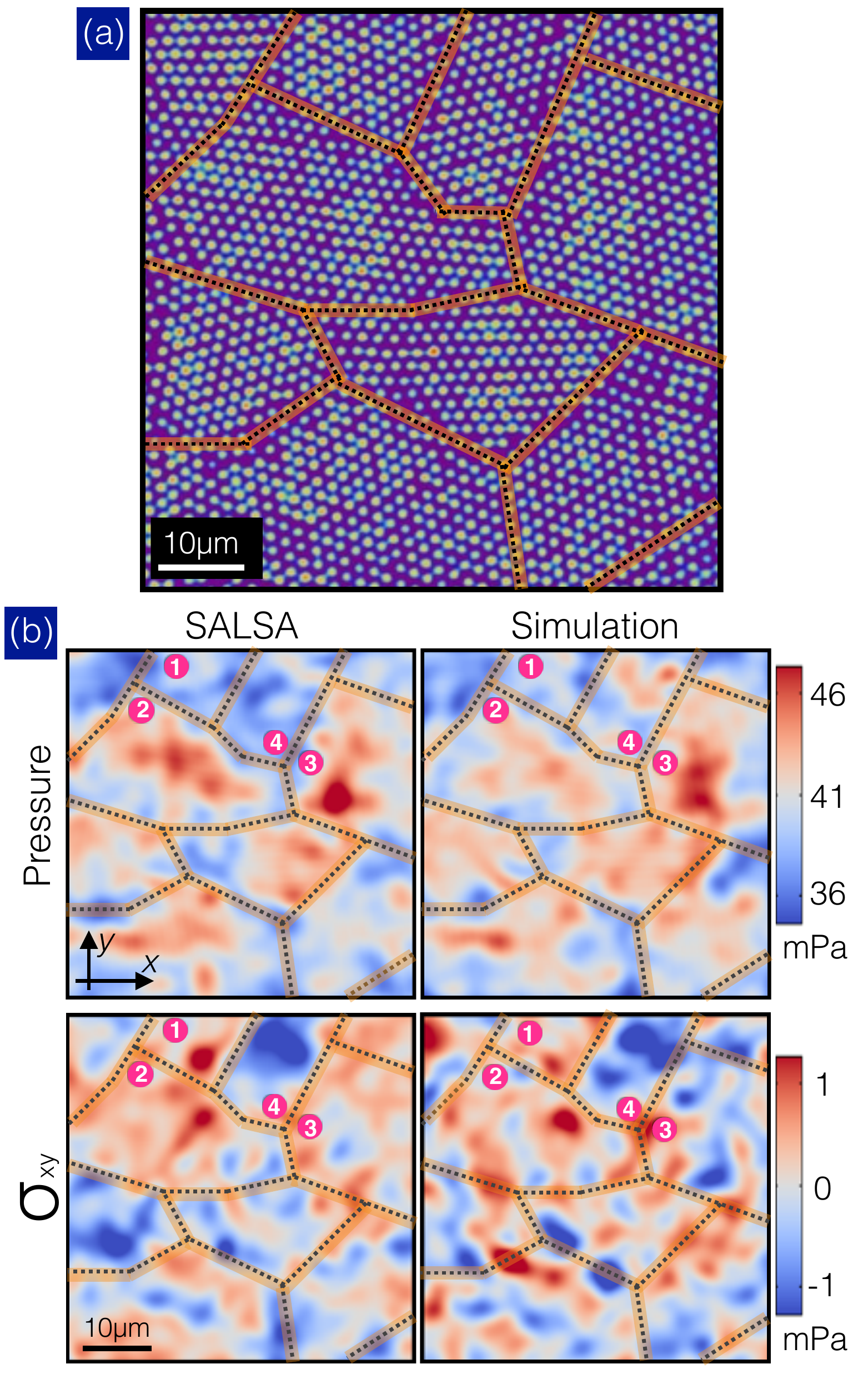}
\caption{\textbf{Stress near grain boundaries.} (a) One confocal image slice of a time series consisting of fifty 3D stacks. (b) Pressure (upper) and $\sigma_{xy}$ (lower) fields of the polycrystal. Both experimental (left) and simulation (right) results show qualitatively similar stress distribution features. The grain boundaries are indicated with dashed lines.}
\label{polycrystal}
\end{figure}

{\bf\emph{Grain boundaries}} are 2D structures important for crystal growth~\cite{friesen2002reversible}, melting kinetics~\cite{alsayed2005premelting, peng2010melting}, transport properties~\cite{gokhale2012directional}, and can substantially harden materials through internal stress variation~\cite{robinson2009coherent, king2008observations, levine2006x}. While X-ray microbeam experiments have been used to reveal {\it strain} fluctuations at the scale of $100~\rm{nm}$~\cite{levine2006x}, measuring {\it stress} remains challenging at these scales, especially at the grain boundaries where particles are highly disordered. 

To visualize such stresses using SALSA, we grow polycrystals using the same method described in the vacancy section (see Fig.~\ref{polycrystal} (a) for a confocal image). We plot the measured pressure and shear stress $\sigma_{xy}$ in the left column of Fig.~\ref{polycrystal}(b). Just as for the dislocation simulation, we employ the featured particle positions as initial configurations, and simulate stresses in the polycrystal. The simulation results (right column in Fig.~\ref{polycrystal}(b)) show similar features to the SALSA stress distributions in both pressure and shear components. 

The spatial fluctuations in both pressure and shear stress seen in Fig.~\ref{polycrystal} (b) are significant compared to relevant stress scales. The standard deviation in pressure ($\approx$ 6 mPa) is about 15\% of the mean pressure whereas the shear stress fluctuation ($\approx 0.7$ mPa). To provide intuition, this stress level is about 30\% of the stress magnitude one lattice constant from a dislocation core, the principle component of a tilt grain boundary. Moreover, we find that both pressure and shear stresses fluctuate between and within grains. For example, the mean pressure difference between grain 1 and 2 is 5 mPa (25\%) whereas grain 3 shows an {\it intragrain} fluctuation of $\approx$10\% the mean pressure. Similar trends can be seen in the shear stress difference between grains 1 and 4, and the fluctuations within grain 3.

Overall, our observation of the stress fluctuations in the polycrystal is consistent with previous simulations~\cite{schiotz1998softening}, and X-ray microbeam measurements~\cite{king2008observations, levine2006x}, where neighboring grains consisting of millions of atoms were found to have substantially different strains. The SALSA measurements indicate such stress fluctuations also arise {\it within} grains consisting of only hundreds of particles. These small crystallites are reminiscent of the nano-scale grains in atomic crystals. Previous atomistic simulations have predicted the stress fluctuations in a strained nanocrystal are predominately localized to the grain boundaries~\cite{schiotz1998softening}. In our colloidal crystal grains however, the stress fluctuations are spread roughly evenly throughout the grains (See SI for direct comparison). Our sample however has not been subject to shear. We conjecture that condensation of stress under plastic strain arises from trapping of dislocations at grain boundaries, grain boundary slip~\cite{schiotz1998softening}, or an as of yet unidentified mechanism.

In conclusion, we measure, for the first time, the microscale stress fields of crystalline defect cores that determine fundamental mechanisms governing processes ranging from local defect interactions to macroscale yielding. We illustrate the specific significance of this microscale measurement in three canonical defects. The measured pressure enhancement around the vacancy core settles the controversy between theory and simulations, and provides critical insights into the origin of attraction between vacancies. The observed softening at the dislocation core validates the decades-long conjecture of the non-singular stresses in numerous dislocation simulations. Finally, the evenly-spread stress fluctuation in the polycrystal predicts hardening of grains when the crystal undergoes plastic deformation. Such stress measurements will be even more valuable when applied to systems driven further out of equilibrium by applied strains since it will directly measure the stress precursors that generate material failure.

\end{document}